\documentclass[conference]{IEEEtran}
\IEEEoverridecommandlockouts

\usepackage{multirow}
\usepackage{cite}
\usepackage{caption}
\usepackage{amsmath,amssymb,amsfonts}
\usepackage{graphicx}
\usepackage{textcomp}
\usepackage{xcolor}
\usepackage{amsmath}
\usepackage{algpseudocode}
\usepackage{algorithm}
\usepackage{subfig}
\usepackage[english]{babel}
\usepackage[utf8]{inputenc}

\usepackage[nottoc]{tocbibind}
\algnewcommand\algorithmicforeach{\textbf{for each}}
\algdef{S}[FOR]{ForEach}[1]{\algorithmicforeach\ #1\ \algorithmicdo}
\usepackage{algorithm}
\usepackage{eso-pic}

\author{
    \IEEEauthorblockN{Ramin Hasibi\IEEEauthorrefmark{1}, Matin Shokri\IEEEauthorrefmark{2}, and Mehdi Dehghan Takht Fooladi\IEEEauthorrefmark{1}}
    \IEEEauthorblockA{\IEEEauthorrefmark{1}Department of Computer Engineering and Information Technology \\
    Amirkabir University of Technology 
    \\\{r.hasibi.94, dehghan\}@aut.ac.ir}
    \IEEEauthorblockA{\IEEEauthorrefmark{2}Department  of  Electrical  and Computer Engineering\\
    K. N. Toosi University of Technology
    \\\{shokri\}@email.kntu.ac.ir}
}
\begin{document}
\captionsetup[figure]{font={small},labelformat={default},labelsep=period,name={Fig.},}
\title{Augmentation Scheme for Dealing with Imbalanced Network Traffic Classification Using Deep Learning}

\maketitle
%Department of Computer Engineering and Information Technology \\
%Amirkabir University of Technology \\
%Email: {\{moein.e, shiva.kt, rasti\}@aut.ac.ir}}

%}\\
%
%,~\IEEEmembership{Senior Member,~IEEE}
%    \thanks{Manuscript received ?. This work is supported in part by Amirkabir University of Technology, Tehran, Iran, and in part by ?.}% <-this % stops a space
%    %\thanks{M. Rasti is with the Department of Electronic and Computer Engineering, Shiraz University of Technology, Shiraz, Iran. Prior to this, he was with the Department of Electrical and Computer Engineering, Tarbiat Modares Univesity, Tehran, Iran.}
%\thanks{Ekram Hossain is with.}% <-this % stops a space
%         }%
%    \markboth{IEEE Transactions on Wireless Communications}{}
    \maketitle

\begin{abstract}
 One of the most important tasks in network management is identifying different types of traffic flows. As a result, a type of management service, called Network Traffic Classifier (NTC),  has been introduced. % There are several types of NTCs that help the network administrator in classifying the applications in the network traffic.
  One type of NTCs that has gained huge attention in recent years applies deep learning on packets in order to classify flows. Internet is an imbalanced environment i.e., some classes of applications are a lot more populated than others e.g., HTTP. Additionally, one of the challenges in deep learning methods is that they do not perform well in imbalanced environments in terms of evaluation metrics such as precision, recall, and $\mathrm{F_1}$ measure. In order to solve this problem, we recommend the use of augmentation methods to balance the dataset. %Augmentation is a method that tries to generate more data through ways that will not change the nature of a class such as flipping a picture horizontally or trying to synthesize artificial data from a certain class. 
  In this paper, we propose a novel data augmentation approach based on the use of Long Short Term Memory (LSTM) networks for generating traffic flow patterns and Kernel Density Estimation (KDE) for replicating the numerical features of each class. First, we use the LSTM network in order to learn and generate the sequence of packets in a flow for classes with less population. Then, we complete the features of the sequence with generating random values based on the distribution of a certain feature, which will be estimated using KDE. Finally, we compare the training of a Convolutional Recurrent Neural Network (CRNN) in large-scale imbalanced, sampled, and augmented datasets. The contribution of our augmentation scheme is then evaluated on all of the datasets through measurements of precision, recall, and $\mathrm{F_1}$ measure for every class of application. The results demonstrate that our scheme is well suited for network traffic flow datasets and improves the performance of deep learning algorithms when it comes to above-mentioned metrics.
\end{abstract}
%\smallskip
%\noindent
\vspace{-2mm}
%\textbf{\emph{Keywords}---}\textbf{Long-Term\mbox{ }Evolution;\mbox{ }Random\mbox{ }Access Overload; Machine-to-Machine Communication; Random Access Channel; Internet of Things}

\section{Introduction}
\label{I. Introduction}
With the ever-increasing amount of traffic that goes through the network, network management has become a difficult task. One of the most important tasks in network management is identifying the types of traffic that are  passing  through  the  network. Classifying the applications is a fairly simple task with high  evaluation metrics. Additionally, NTCs have been able to take care of this matter efficiently. Two major purposes of NTCs are detecting anomalies in the network and classification of applications for Quality of Service (QOS) purposes\cite{maghalebasecrnn,deepbeliefintrusion}. There have  been several types of NTCs that use different methods for handling the task at hand, however, each one has its own drawbacks. These methods are generally divided into three categories as follows
\begin{itemize}
    \item \textbf{Port-based:} This approach is not efficient since some applications do not use a specific port e.g.,  Bittorrent. Moreover, if the port is changed, this method is no longer reliable\cite{maghalebasecrnn}.
    \item \textbf{Deep Packet Inspection (DPI):} These applications use the patterns in the payloads of packets for classification. They generally have three major drawbacks. The first one is that they need to be updated with new patterns in the payloads of emerging applications \cite{maghalebasecrnn}. In addition, they are not able to  identify  all  of  the  flows. Furthermore, if we do not have access to the payload of packets for privacy reasons, their accuracy is very much affected.
    \item \textbf{Machine Learning based:} The  flaws of the  two above  methods  has  gained  attention  to the third type of classifiers, which use machine learning  and specifically deep learning algorithms. This type of algorithms usually work with the features in the header of the packets, but some of them may also take into account the information in the payloads\cite{maghalebasecrnn,sharifia}. Although they are still limited, they have shown great potential in terms of evaluation metrics and will be a great substitution  in  the future for the aforementioned methods.
\end{itemize}

Most of the traces that are gathered from real Internet traffic are imbalanced i.e., some types of application  flows  are  generally  more populated than others  e.g.,  HTTP\cite{elsevierimbalancedgravity,acganaug,devide&conqueraug,Shafiq2018}. This matter is a lot bolder when it comes to large-scale traffic and will cause some serious problems in the way of algorithms' $\mathrm{F_1}$ measure.

Augmentation is an approach in machine learning that addresses the issue of small amount of data for training. This approach usually tries to increase the training data in a way that can be still classified in the same category. Augmentation is a popular method used especially in image classification \cite{effectiveaugimage} and can be done through methods like cropping, zooming, rotating, and filliping vertically or horizontally. Another way of achieving augmentation is through generating artificial data for a class.

In order to address the challenges of machine learning algorithms in imbalanced network datasets, we introduce a novel augmentation method to improve the accuracy of deep learning algorithms on real-world traffic traces by using KDE and LSTM.

The remainder of this paper is organized as follows: In section II we review the related works in the area of NTCs. In section III we describe our augmentation scheme. The dataset and deep learning model that was used in order to classify the traffic traces are mentioned in sections IV and V, respectively. Finally, the evaluation of our method is demonstrated in section VI.
\section{related works}
Due to the high variety in the classes, datasets, and
performance metrics that are used in this area, having a
comparison between the works in this subject is a difficult task \cite{maghalebasecrnn}.
Considering this, there are several well-known pieces of research that
are done up to this point.

There are several works that have applied deep learning
architectures or neural networks in order to solve the classification problem. In \cite{maghalebasecrnn} Lopez-Martin et al. have presented a deep
Convolutional Recurrent Neural Network architecture in order
to classify network flows and have found the best setting in that
environment in terms of hyper-parameters and feature set. Nevertheless,
they have not taken any measures to handle the imbalance
problem of their dataset. Additionally, the scale of their dataset is approximately fifth of the one that we are using.

In \cite{springer_deep} Rahul et al. have also  proposed using Convolutional Neural Networks (CNNs) in
order to classify network traffic but only consider three classes
of applications in their work on a limited amount of data.

In \cite{sharifia} a comparison between CNN and Stacked Auto-Encoders in order to classify not only types of traffic but also applications in the network in a standard VPN/none-VPN dataset in packet level has been given.
The scheme that was used, unlike ours, relies on the features from both
header and payload of packets which may not be available in
some privacy-preserving datasets.

Finally, in \cite{bayesianneural} Auld et al.
have deploy a Bayesian neural network in the form of a multi-layer
perception and accordingly classify their dataset. In
this work, the lowest performance metrics are from the classes
with the lowest number of data. 

Some works in this area have attempted to battle the imbalanced property through different measures. In \cite{PGMMOORE} Rotsos et al. have
introduced a method through Probabilistic Graphical
Models for semi-supervised learning in a Naive Bayes model.
For their learning, they have assumed a Dirichlet distribution
prior for the classes with high $\alpha$ value. This is based on the assumption that some classes  have a higher probability than others. 

In addition, in \cite{acganaug}, an augmentation method has been proposed by using
an Auxiliary Classifier Generative Adversarial
Network (AC-GAN), although only two classes of network is considered: SSH and none-SSH. Furthermore, their method is only evaluated on traditional machine learning algorithms like Support Vector Machines, Random Forest,
and Naive Bayes.
Also \cite{devide&conqueraug}  has presented a new feature extraction
method using a divide and conquer approach for an imbalanced
dataset in the network.

As an instance of LSTM used for generating sequential data, \cite{LSTMGEN} has introduced a method to generate data using LSTM and evaluated the method to show that it can capture the temporal features in the dataset. 
LSTM has also been used as an augmentation tool in works such as \cite{daskhataug} and \cite{LSTM_ae_aug} for generating handwriting and human movement data, respectively, and has proven to be efficient in both cases.

\section{augmentation scheme for generating time series network data}
In this section, we describe our contribution of augmentation scheme for generating new data in network traffic traces. 

Every flow in the network has the same 5-tuple attributes: 
\begin{itemize}
    \item Source IP address.
    \item Destination IP address.
    \item Source port number.
    \item Destination port number.
    \item Link layer protocol e.g., TCP and UDP.
\end{itemize}
Every application in the internet creates a flow of packets between communicating peers \cite{naivebayesaggregate}.

In order to represent flows in our work, we have to choose a set of features for each one that can capture the nature of a flow. According to \cite{maghalebasecrnn}, the appropriate set of features that will give acceptable results for classifying flows are mentioned in Table \ref{table:featuretable}. These features are gathered for the first 20 packets of each flow, which are more than enough for capturing the temporal and spatial features of a flow.

\begin{table}[t!]
\caption{Features of each flow}
\label{table:featuretable}
\begin{center}
\begin{tabular}{ |p{3cm}||p{1.25cm}| } 
 \hline
 \textbf{Feature} & \textbf{Type} \\
 \hline
Source port & \multirow{4}{4em}{Numerical} \\
Destination port & \\
 Inter-arrival time &\\
 Payload length &\\
 \hline
Direction of packet & \multirow{2}{4em}{Sequential} \\ 
TCP window size & \\
\hline
\end{tabular}
\end{center}
\end{table}

As shown in Table \ref{table:featuretable}, we can put the features in two categories: sequential and numerical. Each group has its own way of augmentation which are described in the following.
\subsection{Generating sequential features}

In this section, we demonstrate our approach to generating sequential features.

As mentioned earlier, traffic flow comprises the sequence of packets that are transmitted between a source and a destination. Some applications are uni-directional i.e., the packets are only transmitted in one direction e.g., uploading a data. However, in some type of applications, packets go in both directions such as when a client is communicating with the server and gets a response for its request. Whether a packet is sent from source or destination depends on the sequence of packets that have already been sent up to this point in the flow.  Therefore, we can conclude that the sequence of directions of packets in a specific application is of time-series nature and can be generated through means of sequence generation like in \cite{daskhataug,LSTM_ae_aug}.
TCP window size is another feature of the flow that is dependent on the previous values in the flow. Generally, this value is an indicator of the conditions of the connection and processing speed of data in the flow \cite{rfc1323}. Thus, its amount at each step of the flow is affected by previous steps' values.

One of the most common ways to generate a sequence is using Recurrent Neural Networks (RNNs), which try to learn the patterns in time-series data e.g., speech, music, text, etc\cite{RNN}. In our work, we use one type of RNNs called LSTM networks\cite{LSTM}. Each LSTM block tries to learn the probability distribution in a step of a sequence whilst taking into consideration the information from previous steps. 

In order to train the network, we gathered the patterns of packet directions in a flow for up to 20 packets in a class of flow application. We encode every direction by 1 or 0 with the former being from source to destination and the latter is the other way around. At the end of each sequence, we put a unique character as an indicator of the ending of the flow. Then every sequence is shifted by one character to the right and is used as labels in order to train each step of generation in LSTM.

\begin{figure}[t!]
\centering
\includegraphics[width=7.70cm, height=6cm]{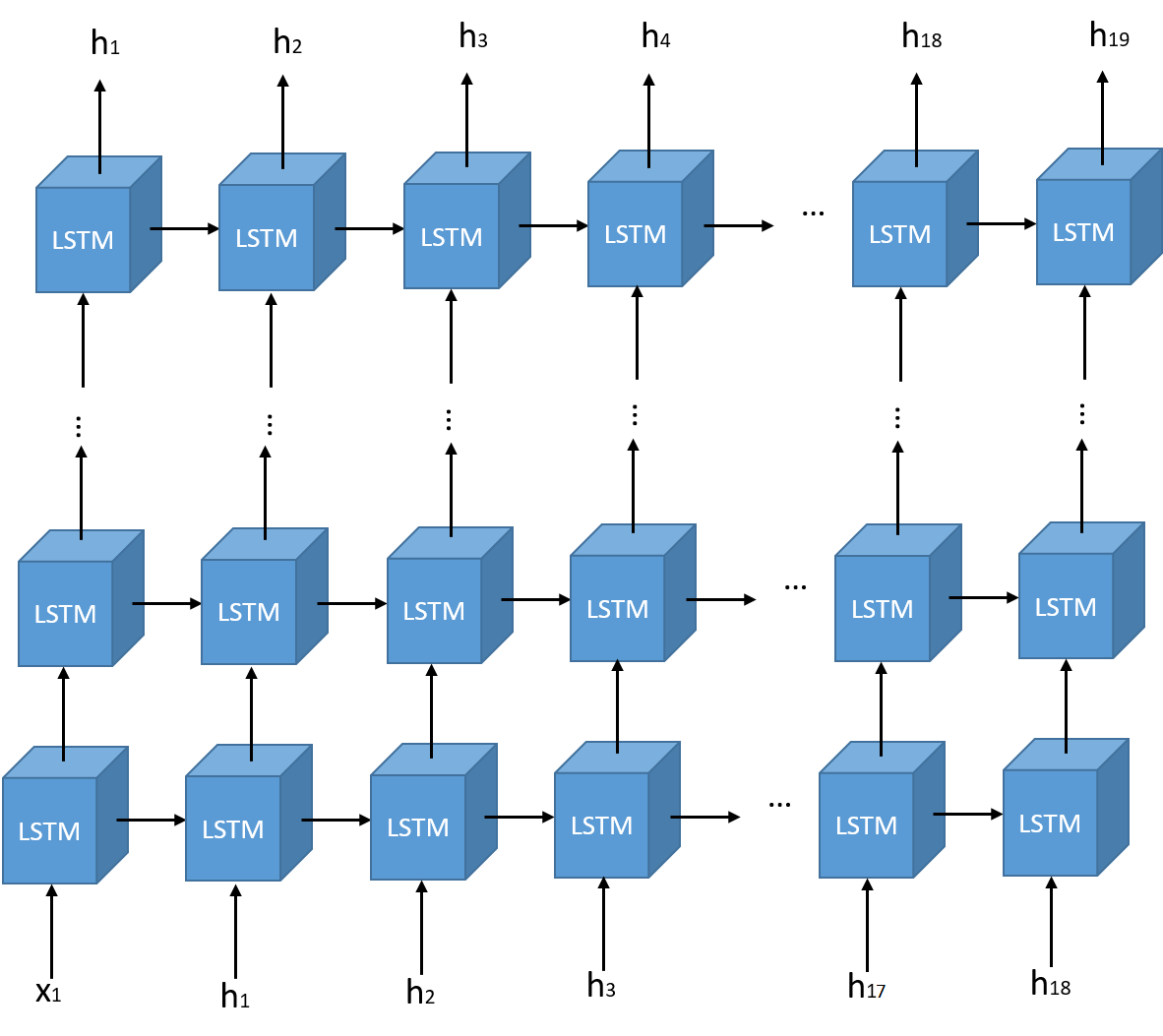}
\caption{Generating 19 time steps of packet sequence with LSTM}
\label{fig:LSTM}
\end{figure}

In the generation phase, first, we choose a direction based on the distribution of that direction in the dataset for the first time step and give that as input to the LSTM. Afterwards, we use the output of each step as probability distribution of each character (1, 0, or ending character) and generate a new direction. Then, we feed that output direction to LSTM in order to generate next step probabilities. The maximum number of steps are 19 in order to generate the pattern of flow up to 20 packets (first packet is always from source to destination). Let $x_t$ and $h_t$ denote the direction of the packet  in the dataset and the generated direction by the LSTM at time step $t$, respectively. Therefore, the generation process is demonstrated in Fig. \ref{fig:LSTM}.

In order to  generate window size values, we use the same scheme, although the characters in this case are the values of window sizes in our dataset instead of 0 and 1.

\subsection{Generating numerical features}

In this section, we describe our method of generating numerical features of a flow.

As shown in Table \ref{table:featuretable}, we consider four numerical features for each packet of the flow. In order to generate new samples from these features, first, we need to learn their probability distribution. Since these features are not sequential, we can use conventional probability density distribution estimation methods. One of these methods is KDE that is in the category of kernel methods. 

KDE, also known as the Parzen–Rosenblatt window, is one of the most famous methods used to estimate the probability density function of a dataset. KDE, as a non-parametric density estimator, does not have any assumptions about the density function as opposed to the parametric family of algorithms. This method will learn the shape of the density from the data automatically. This flexibility that arises from its non-parametric nature, makes KDE a very popular method for data drawn from a complicated distribution.

Let $X=\{x_1,x_2,...,x_n\}$ denote the set of independent and identically distributed random samples from a group of features e.g., inter-arrival time and $K(x):{\rm I\!R}^d \mapsto{\rm I\!R}$ denote the probability distribution function (PDF) of the kernel of our choosing. Then we can estimate the PDF of $X$ by

\begin{equation}
    \hat{p} = \frac{1}{nh^d} \sum_{i=1}^{n}K(\frac{x-x_i}{h}),
\end{equation}
where $\hat{p}$ is the estimated PDF of $X$ and $h>0$ is the bandwidth of the kernel that is used to control the smoothing degree of the kernel\cite{Rosenblatt,parzen1962}.

%For estimating the inter-arrival times of packets we choose the Pareto distribution as our kernel. The PDF of kernel in this case is as follows:

%\begin{equation}
%    K(x) = \frac{ab^a}{x^{a+1}}
%\end{equation}

%The reasoning behind this choice of kernel is that just like Pareto distribution the probability of inter-arrival decreases as the value of time is increased i.e., it is less possible for a packet to be transmitted in a flow a long time after the previous packet.

%In order to generate payload length we use normal distribution kernel with the formula:
The common examples of K(x) is Gaussian distribution with $\mu = 0$ and $\sigma = 1$ as expressed by
\begin{equation}
    K(x) = \frac{1}{{\sqrt {2\pi } }}e^{{{ - \left( {x  } \right)^2 } \mathord{\left/ {\vphantom {{ - \left( {x - \mu } \right)^2 } {2\sigma ^2 }}} \right. \kern-\nulldelimiterspace} {2}}},
\end{equation}
which is also used for our scheme.

In order to prevent bias-variance problem in fixed $h$ cases, we used the bandwidth selection method represented in \cite{silverman}. According to \cite{silverman} if a Gaussian kernel is used, it can be shown that the optimal value of $h$ is $h^*=\Big(\frac{4\hat{\sigma}^5}{3n} \Big)^\frac{1}{5}$.

%Where $\sigma^2$ and $\mu$ denote the standard deviation and average of the kernel, respectively.

\section{dataset}
\label{datasetsec}
In this section, we describe our dataset and its labeling method.

For this paper, we used real traces of traffic from the campus of Amirkabir University of Technology that includes more than 70 gigabytes of packets from UDP and TCP link layer protocols. Next, we label flows using nDPI, which is an open source DPI tool released by ntop for classifying the flows based on applications\cite{nDPI}. The reason for our choice of labeling tool is that according to \cite{NDPICOMP}, nDPI is the most accurate open-source DPI tool among available DPI tools. 

Nineteen classes of traffic from more than 50 gigabytes of packets were chosen which include 904490 flows. 85 percent of these flows were chosen for training and the rest are used for test dataset. The classes of applications are the ones with the most number of instances in the dataset and can be seen in Table \ref{table:classes}.

\begin{table}[t!]
\caption{Classes of applications}
\label{table:classes}
\begin{center}
 \begin{tabular}{||c c||} 
 \hline
\textbf{Class} & \textbf{Number of Flows}\\
\hline
\hline
HTTP & 58774 \\
DNS & 126960 \\
NTP & 4633 \\
BitTorrent & 6146 \\
HTTP\_Download & 16326 \\
SSL\_No\_Cert & 10603 \\
Steam & 4460 \\
RDP & 1425 \\
SSL & 341846 \\
SSH & 9746 \\
Facebook & 2772 \\
Twitter & 2198 \\
Google & 96072 \\
WindowsUpdate & 2343 \\
Telegram & 186256 \\
Instagram & 6683 \\
Microsoft & 18196 \\
PlayStore & 5304 \\
YouTube & 3747 \\
 \hline

\end{tabular}
\label{tab1:population}
\end{center}
\end{table}

\begin{figure}[t!]
\centerline{\includegraphics[width=9.70cm, height=8cm]{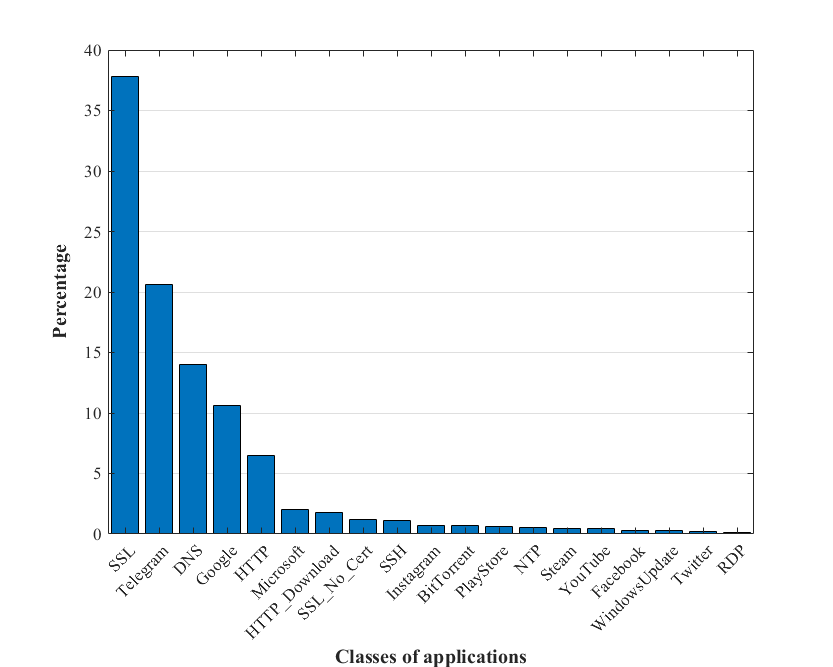}}
\caption{the percentage of different classes of applications in our dataset.}
\label{fig:populationpercent}
\end{figure}

As shown in Table \ref{tab1:population}, there are different classes of applications in our dataset and the names of our labels are chosen based on the labels given by nDPI. The percentage of each class is shown in Fig. \ref{fig:populationpercent}.

As demonstrated by the bar chart, the imbalance feature of the dataset is clear. The most populated class of appliaction is SSL with more than 37 percent of the population and the least populated class is RDP with less than 0.16 percent. Furthermore, more than 83 percent of the whole dataset consists of only 4 classes. Additionally, 10 classes have less than 1 percent, which are the less populated classes and therefore, some of them are expected to be susceptible to low evaluation metrics.

\section{classification scheme}

In this section we explain the classification scheme that was used to test our augmentation.

The classification process mainly consists of two stages:
\begin{itemize}
    \item Augmentation phase.
    \item Training phase.
\end{itemize}
\subsection{Augmentation} 
\label{Augmentation}
In the augmentation phase, we  generate new data from classes that have less population in the dataset. First, we train and use LSTM to generate the pattern of directions and TCP windows sizes in the flow. After that, we estimate the PDFs of every single numerical feature using KDE. Then, according to these PDFs, we generate points in every feature domain. These points are our generated features for the packets. Finally, we generate up to 20 packets per flow and put these features in an array of size 6*20 (6 features from 20 packets). If the number of packets in the generated sequence is less than 20, the rest of the array is appended with 0. These arrays will comprise the generated dataset.   %and then build the Cumulative Distribution Function (CDF) from the PDF. Then, according to normal distribution we choose random points between 0 and 1 and find the corresponding point from the feature domain based on CDF. 

The pseudo-code for the augmentation process is given in Algorithm \ref{alg:the_alg}.

\begin{algorithm}[t!] 
\caption{Augmentation process}
\label{alg:the_alg}
\hspace*{\algorithmicindent} \textbf{Input} set $C$ of classes with low population \\
\hspace*{\algorithmicindent} \textbf{Output} set $\hat{C}$ of generated flows
\begin{algorithmic}[1]
\ForEach {$c \in \mathcal C $}
\State $p_d \gets $ pattern of directions in $c$
\State $p_{tcp} \gets $ pattern of TCP windows in $c$ 
\State Train LSTMs for each pattern in $p$ and $p_{tcp}$
\State $\hat{p}_d \gets $ generated direction patterns from  LSTM
\State $\hat{p}_{tcp} \gets $ generated  TCP window patterns from  LSTM
\State $\hat{p} \gets \hat{p}_d \cup \hat{p}_{tcp}$
\State $NF \gets $ sets of Numerical features in $c$
\ForEach {set $nf \in  NF$}
\State  $PDF(nf)\approx KDE(nf)$
\State  $rs \gets $ generated random samples from $PDF(nf)$
\EndFor
\State $gen\_flows \gets$ new data from $rs$ and $\hat{p}$ based on \cite{maghalebasecrnn}
\State $\hat{C} \gets \hat{C} \cup gen\_flows$
\EndFor
\end{algorithmic}
\end{algorithm}

\subsection{Training}
Next, we train a Convolutional Recurrent Neural Network on the augmented dataset. In order to do this, we choose the architecture that was suggested in \cite{maghalebasecrnn}. This architecture includes two Convolution layers, the sizes of which are 32*4*2 and 64*4*2, respectively. Each of these layers is followed by a Batch Normalization (BN) layer. After that, the output of the last BN is put in time-series format and is fed into an LSTM layer of 100 hidden units. At the end of the architecture, there are two Fully-Connected (FC) layers, each with 100 and 108 hidden nodes and dropout rates of 0.2 and 0.4, respectively. These are followed by a soft-max layer with 19 outputs, each corresponding to 19 classes of traffic. The activation function of every layer in this architecture, except for the soft-max layer, is Relu function.
\section{evaluation}
In this section, we present the evaluation results of the model on three different datasets.

In order to fully discover the advantages of our method, three sets of datasets are prepared: 
\begin{itemize}
    \item Actual data: The exact dataset from section  \ref{datasetsec}.
    \item Sampled data: Dataset of section \ref{datasetsec} over-sampled using \cite{Sampling}.
    \item Augmented data: Dataset of section \ref{datasetsec} augmented using our method.
\end{itemize}
The method of sampling in \cite{Sampling} is a simple yet effective approach to handle the problem of imbalanced classification and is widely used in many works such as in \cite{sharifia}.

Classes NTP, Facebook, twitter, WindowsUpdate, Instagram, PlayStore, and YouTube are chosen for augmentation and over-sampling because the CRNN network gets the worst results in these classes. Furthermore, these are the classes that have low number of samples in the dataset.

The evaluation metrics that are chosen to measure the performance of our approach are those that are mostly used for imbalanced datasets and give an appropriate analysis of the methods that are employed. These metrics are precision, recall, accuracy, and $\mathrm{F_1}$ measure, whose formulas are given in the following.

\begin{equation}
    Precision = \frac{\text{TP}}{\text{TP}+\text{FP}},
\end{equation}
\begin{figure}[t!]
\centerline{\includegraphics[width=9.70cm, height=8cm]{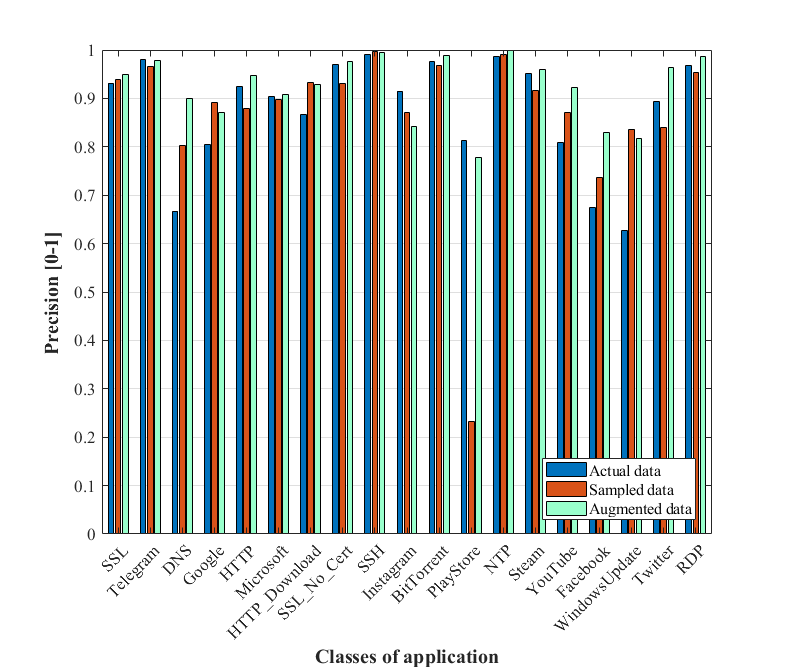}}
\caption{Precision measure comparison of per class in the dataset}
\label{fig:precision}
\end{figure}

\begin{figure}[t!]
\centerline{\includegraphics[width=9.70cm, height=8cm]{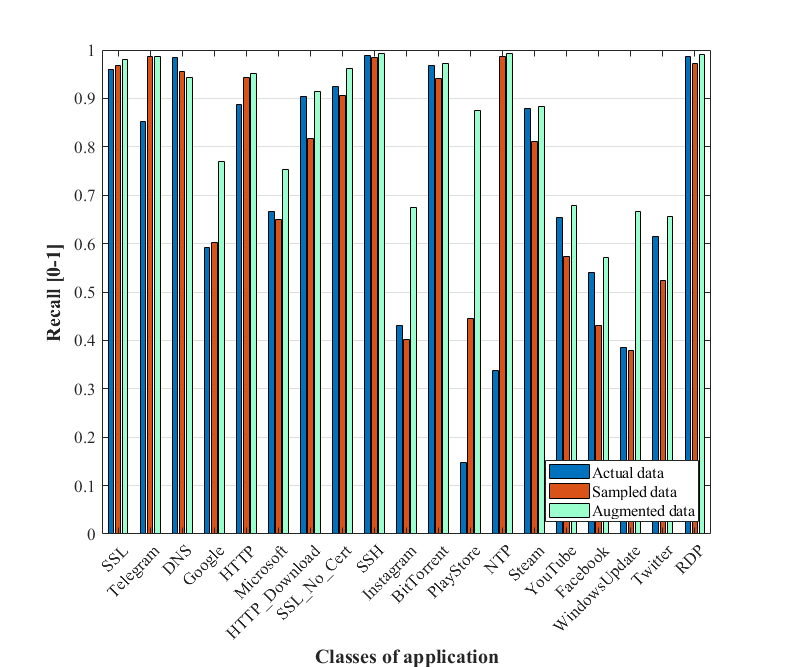}}
\caption{Recall measure comparison of per class in the dataset}
\label{fig:recall}
\end{figure}

\begin{figure}[t!]
\centerline{\includegraphics[width=9.70cm, height=8cm]{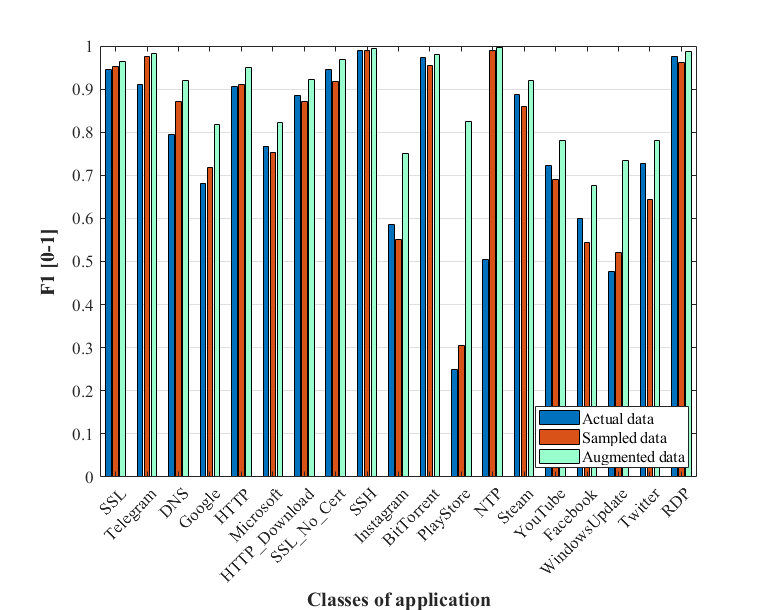}}
\caption{$\mathrm{F_1}$ measure comparison of per class in the dataset}
\label{fig:f1}
\end{figure}
\begin{equation}
    Recall = \frac{\text{TP}}{\text{TP}+\text{FN}},
\end{equation}
\begin{equation}
    Accuracy = \frac{\text{TP}+\text{TN}}{\text{TP}+\text{TN}+\text{FP}+\text{FN}}.
\end{equation}
The TP, FP, TN, and FN in above formulas depict true positive, false positive, true negative, and false negative values, respectively.
\begin{figure}[t!]
\centerline{\includegraphics[width=9.70cm, height=8cm]{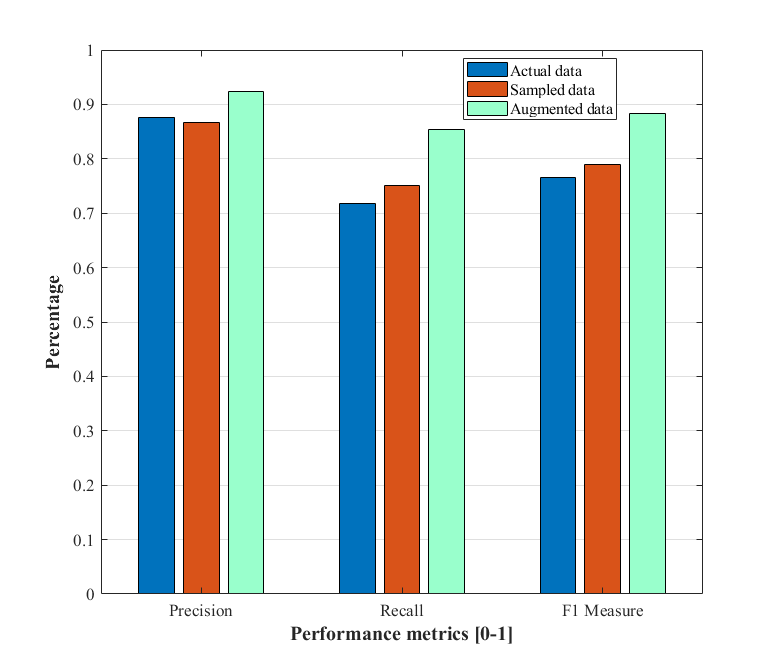}}
\caption{Comparison of Total Precision, Recall, and $\mathrm{F_1}$ measure resulted from both methods}
\label{fig:overall}
\end{figure}

\begin{figure}[t!]
\centerline{\includegraphics[width=9.70cm, height=8cm]{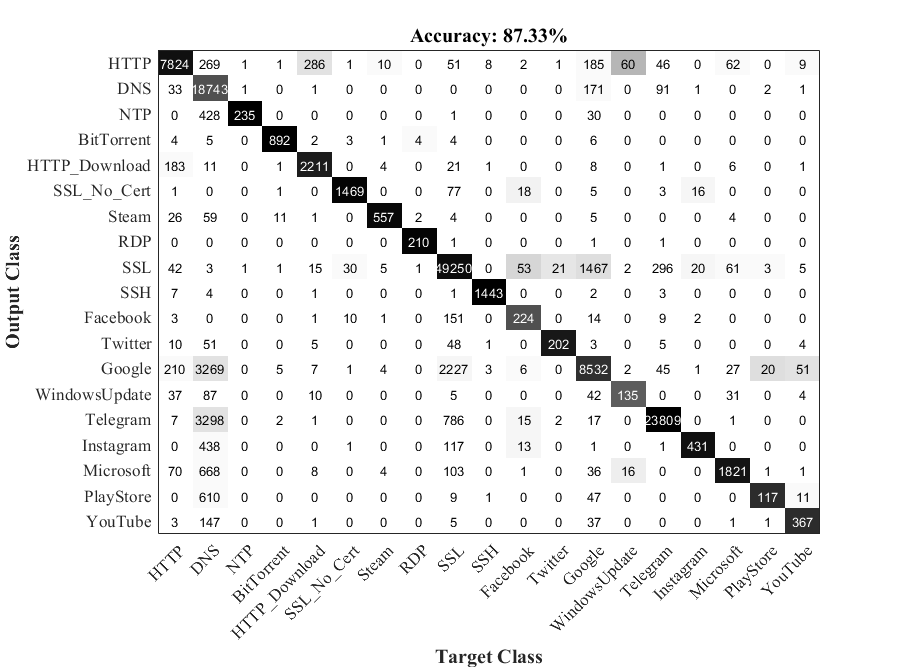}}
\caption{Confusion matrix resulted from the actual dataset}
\label{fig:evaluation:normalConf}
\end{figure}
\begin{figure}[t!]
\centerline{\includegraphics[width=9.70cm, height=8cm]{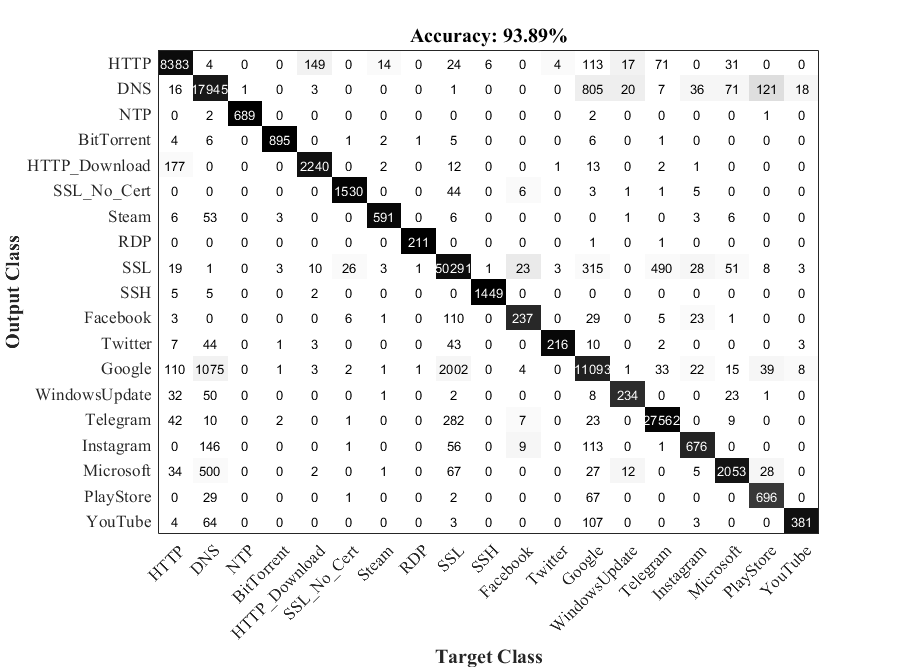}}
\caption{Confusion matrix resulted from the augmented dataset}
\label{fig:evaluation:augmentedConf}
\end{figure}

%The word true (false) in above-mentioned values demonstrates whether the prediction of the algorithm about the class is correct (incorrect). Additionally, the word positive (negative) shows the actual  class of the data. For example, false positive values are those that the algorithm has falsely detected a class other than their actual one.

\begin{equation}
    F_{1}=2*\frac{Precision \times Recall}{Precision + Recall}.
\end{equation}
The $F_1$ measure shows the overall performance of algorithm on both precision and recall.
%In order to demonstrate the performance of our purposed scheme, the sampling technique of \cite{Sampling} ,which has also been performed in \cite{sharifia}, is also done on the data and the results are compared to those of our augmentation approach. 

In Fig. \ref{fig:precision} the precision metric for all three datasets  is given. Although in some classes with less instances that have been augmented like Playstore and Instagram, there has been a slight decrease in precision, others have mostly had an improvement in this matter. In some cases, the sampled dataset performed better than our method such as BitTorrent and Google, but due to the lack of generalization, we can see that in a class like Playstore, which is sampled in large scales, this  method has a huge decrease in the results. Furthermore, the number of classes that are improved by our augmentation is more than those that performed better in sampled dataset.

Fig. \ref{fig:recall} depicts the recall of each class in three separate datasets. In every augmented class, there is a clear upgrade in recall measure. This is due the fact that the number of FN predictions are less for these classes compared to the normal dataset. This might have some negative effect on the over-populated class of DNS, but for others this metric is improved. Due to higher generality in our augmentation, it is obvious that the amount of increase in recall in our approach is  higher than sampling in augmented classes in every instance. Moreover, sampling has caused a decrease in recall in 12 classes compared to actual dataset.

Fig. \ref{fig:f1} illustrates the $\mathrm{F_1}$ measure in all the classes of the dataset.  This figure verifies the fact that overall performance of our method is better than sampling in each and every one of the classes.

Fig. \ref{fig:overall} shows the overall measures on the whole datasets. As shown in this figure, although sampling improved the recall, it has also a slight decrease in precision due to the lack of generalization. However, the overall performance as shown by the increase, albeit a small one, on the $\mathrm{F1}$ is better than the actual dataset. On the other hand in our method, in all three metrics, there is a noticeable improvement which is more than any that is caused by sampling method. 

Fig. \ref{fig:evaluation:normalConf} and Fig. \ref{fig:evaluation:augmentedConf} illustrate the confusion matrices of actual and augmented datasets, respectively. As shown in Fig. \ref{fig:evaluation:normalConf} classes of HTTP, DNS, and SSL, which have high number of instances in the dataset, have noticeable negative effect on majority of classes' prediction. Fig. \ref{fig:evaluation:augmentedConf} shows that our method is able to improve this matter and lessen the number of false predictions. Additionally, the number of true positives in HTTP and SSL is increased. Although DNS predictions have less true positives, the number of false negatives is diminished. Moreover, the overall accuracy in our method is increased by 6.56 percent.

\section{conclusion}
In this paper, we proposed an augmentation method for imbalanced network traffic classification on real traffic traces based on LSTM and KDE. In order to compare the performance of our scheme, we considered two sampled and augmented datasets. The results that are obtained from CRNN show that our approach gets better results in overall measures of precision, recall, and $\mathrm{F_1}$.

\end{document}